\documentclass[]{BioPhysMath}
\usepackage[OT4,T1]{fontenc}
\usepackage[utf8]{inputenc}
\usepackage{polski}
\usepackage{graphicx}

\usepackage{amsmath}
\usepackage{amssymb}

\usepackage[english,polish]{babel} 

\usepackage{color}
\usepackage{tabularx}
\RequirePackage[numbers,sort&compress]{natbib}

\usepackage[colorlinks=true]{hyperref}
  \hypersetup{
    pdftitle={Template BioPhysMath}, 
    pdfauthor=Nowy Autor, 
    colorlinks,
    urlcolor=blue,
    filecolor=magenta,
    citecolor=green,
    linkbordercolor={1 1 1}, 
    citebordercolor={1 1 1},  
    urlbordercolor={ 1 1 1}  
  }

\newcommand{\orcid}[1]{\href{https://orcid.org/#1}{\includegraphics[scale=.05]{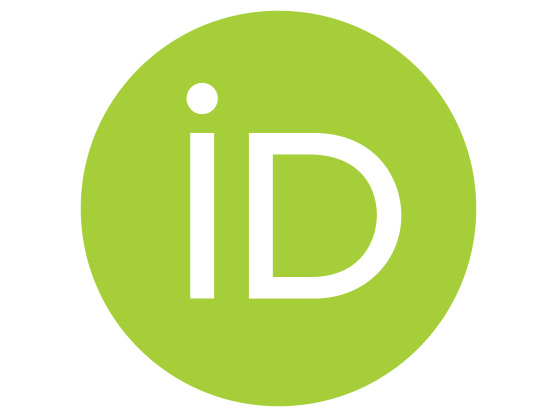}}}

\newcommand{\Cov}{\mathrm{Cov}}

\usepackage{algorithm,algpseudocode}

\newcommand{\rr}{\mathbb R}
\newcommand{\abs}[1]{\left\vert#1\right\vert}
\pdfoutput=1
\usepackage{graphicx}
\usepackage{wrapfig}
\graphicspath{{./Figures/},{./Pictures/}}
\usepackage{multicol}
\firstpage{i}    
\LogoG{\vspace*{1cm}\includegraphics[width=0.1\textwidth]{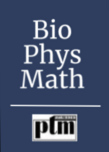}}
\volume{1}
\fasc{???}
\years{2025}
\LogoGMS{\includegraphics[width=0.18\textwidth]{bpm.png}}
\wwwtrue

\pages{23}
\received[13th August 2025]{17th April 2025}
\lastrevision{}
\logo{}{}{}

\title[Testing of tempered fractional Brownian motions]{Testing of tempered fractional Brownian motions}

\tpauthortrue 
\author[K. Macioszek]{Katarzyna Macioszek}
\affiliation{Wrocław University of Science and Technology, Faculty of Pure and Applied Mathematics}
\address{Wyb. Wyspianskiego 27, 50-370 Wroclaw, Poland}
\email{katarzyna.macioszek1@gmail.com}

\author[F. Sabzikar]{Farzad Sabzikar\orcid{0000-0003-0086-4776}}
\affiliation{Iowa State University, Department of Statistics}
\address{Ames, IA 50011, United States of America}
\email{sabzikar@iastate.edu}

\author[K. Burnecki]{Krzysztof Burnecki \orcid{0000-0002-1754-4472}}
\affiliation{Wrocław University of Science and Technology, Faculty of Pure and Applied Mathematics}
\address{Wyb. Wyspianskiego 27, 50-370 Wroclaw, Poland}
\email{krzysztof.burnecki@pwr.edu.pl}


\begin{document}
\vspace{-10ex}
\renewcommand{\thefootnote}{}
\footnote{\href{http://creativecommons.org/licenses/by/3.0/}{Licensed under a Creative Commons Attribution License (CC-BY)}}
\setcounter{page}{1} 
\selectlanguage{english}\Polskifalse

\begin{abstract} 
We propose here a testing methodology based on the autocovariance, detrended moving average, and time-averaged mean-squared displacement statistics for tempered fractional Brownian motions (TFBMs) which are related to the notions of semi-long range dependence and transient anomalous diffusion. In this framework, we consider three types of TFBMs: two with a tempering factor incorporated into their moving-average representation, and one with a tempering parameter added to the autocorrelation formula. We illustrate their dynamics with the use of quantile lines. Using the proposed methodology, we provide a comprehensive power analysis of the tests. It appears that the tests allow distinguishing between the tempered processes with different Hurst parameters.
\end{abstract}

\section{Introduction}
\vskip 0.3cm

Gaussian processes play a central role in statistical modeling due to their strong theoretical foundation and practical versatility. A key advantage of Gaussian processes lies in their complete characterisation by second-order statistics, where a zero-mean Gaussian process is fully defined by its covariance matrix. This property, which arises from the central limit theorem, makes Gaussian processes a natural choice for modeling a wide range of stochastic systems. Fractional Brownian motion (FBM) is a Gaussian stochastic process whose increments, termed fractional Gaussian noise (FGN), can exhibit long-range dependence in the sense that the power-law spectral density of FGN blows up near the origin \cite{beran2013long,embrechts2009selfsimilar,samorodnitsky2016stochastic,pipiras2017long}.
FBM has become popular in applications to science and engineering, as it yields a simple tractable model that captures the correlation structure seen in many natural systems \cite{ascione2021fractional,harms2019affine,kolmogorov1940wiener,mandelbrot1983fractal,meerschaert2019stochastic,molz1997fractional}.

Recently, two broad classes of continuous stochastic Gaussian processes, known as the tempered fractional Brownian motion (TFBMI) and the tempered fractional Brownian motion of the second kind (TFBMI\!I), were introduced in \cite{meerschaert2013tempered} and \cite{sabzikar2018tempered}, respectively. A distinct tempering approach, which directly modifies the autocorrelation structure rather than the moving average representation, was introduced in \cite{MolSan} and is referred to as tempered fractional Brownian motion of the third kind (TFBMI\!I\!I). This variation enables TFBMI\!I\!I to model systems with specific decay characteristics in their autocorrelation functions, making it particularly suitable for applications in biological systems and single-particle tracking experiments.

Unlike FBM, TFBMI and TFBMI\!I can be defined for any value of the Hurst parameter $H > 0$. These processes have garnered significant attention across various fields of research. For instance, bifurcation theory has proven to be a valuable tool for analysing qualitative and topological changes in the orbit structure of parameterized dynamical systems. In \cite{zeng2016bifurcation}, a stochastic phenomenological bifurcation of the Langevin equation perturbed by TFBMI was developed, demonstrating that the tempered fractional Ornstein-Uhlenbeck process, which is the solution to the Langevin equation driven by a TFBMI, exhibits diverse and intriguing bifurcation phenomena. The study in \cite{chen2017localization} further explored the properties of TFBMI, including its ergodicity, and derived the corresponding Fokker-Planck equation. They also demonstrated that the mean squared displacement of the tempered fractional Langevin equation transitions from $t^2$ (ballistic diffusion at short times) to $t^{2-2H}$, and then back to $t^2$ (ballistic diffusion at long times). Arbitrage opportunities in the Black-Scholes model driven by TFBMI were examined in \cite{zhang2017arbitrage}. Additionally, \cite{sabzikar2020asymptotic} developed an asymptotic theory for the ordinary least squares estimator of an autoregressive model of order one when the additive error follows a discrete tempered linear process, showing that the limiting results involve TFBMI\!I under specific conditions. An application of tempered fractional Brownian motion to geophysical flows is presented in \cite{meerschaert2014tempered}, where a tempered fractional time series model for turbulence is proposed. The authors seek to improve the modelling of turbulence in geophysical contexts, demonstrating the practical applicability of tempered fractional processes. The papers \cite{boniece2021tempered,boniece2018tempered} focus on wavelet estimation and modelling of geophysical flows using tempered fractional Brownian motion. \cite{mishura2024asymptotic} constructed least-square estimators for the unknown drift parameters within Vasicek models driven by these processes.

While the theoretical properties and practical applications of TFBMs have been extensively studied, there is still a need for rigorous statistical tests to determine whether a given data set follows a tempered fractional Brownian motion structure. This paper addresses this need by introducing a robust statistical testing framework based on quadratic form statistics. Specifically, we employ three well-known statistical measures: the autocovariance function (ACVF), the detrended moving average (DMA), and the time-averaged mean-squared displacement (TAMSD). These measures are chosen for their ability to capture the essential characteristics of TFBMs trajectories across different scales and configurations.

The main contributions of this paper are as follows.
\begin{itemize}
    \item We propose a statistical testing methodology tailored for TFBM processes using quadratic form statistics.
    \item We evaluate the performance of three statistical measures (ACVF, DMA, and TAMSD) across different TFBM types (TFBMI, TFBMI\!I and TFBMI\!I\!I) within a power study.
    \item We assess the sensitivity of the testing framework to changes in the Hurst index and tempering parameter.
\end{itemize}

In Section~\ref{chap:tfbm_intro}, we provide definitions and main properties of all three types of tempered stochastic processes studied. In addition, we present quantile lines for each process with specified parameters to offer more insight into their behaviour. Section~\ref{chap:test_intro} introduces the testing methodology for TFBMs, which relies on the quadratic form representation of the selected statistics. Section~\ref{sec:power_results} presents the results of the introduced testing methodology, with a focus on examining two-sided tests against various alternative hypotheses to explore their performance comprehensively. Section \ref{sec:conclusions} provides a summary of the results.

\section{Preliminaries of the tempered fractional processes}
\label{chap:tfbm_intro}
\vskip 0.3 cm
Let $B=\{B_s, s\in\mathbb{R}\}$ be a two-sided Wiener process, $H>0$, and $\lambda>0$.

\begin{definition}
The stochastic processes $B^{I}_{H,\lambda}=\{B^{I}_{H,\lambda}(t)\}_{t\in\mathbb{R}}$ defined by the Wiener integral
\begin{equation}\label{eq:TFBMdefn}
	B^{I}_{H, \lambda}(t) := \int_{\mathbb{R}}\Big[g_{H,\lambda,t}^{I}(s)\Big]dB_s,
\end{equation}
where
\begin{equation*}
g_{H,\lambda,t}^{I}(s):=\left(t-y\right)_{+}^{H-\frac{1}{2}} e^{\lambda(t-s)_{+}} - (-s)_{+}^{H-\frac{1}{2}} e^{\lambda(-s)_{+}}, s\in\mathbb{R}
\end{equation*}
is called a tempered fractional Brownian motion (TFBMI), see \cite{meerschaert2013tempered}.
\end{definition}

\begin{definition}
The stochastic processes $B^{I\!I}_{H,\lambda}=\{B^{I\!I}_{H,\lambda}(t)\}_{t\in\mathbb{R}}$ defined by the Wiener integral
\begin{equation}\label{eq:TFBMdefnII}
	B^{I\!I}_{H, \lambda}(t) := \int_{\mathbb{R}}\Big[g_{H,\lambda,t}^{I\!I}(s)\Big]dB_s,
\end{equation}
where
\begin{equation*}
\begin{split}
g_{H,\lambda,t}^{I\!I}(s):&=\left(t-s\right)_{+}^{H-\frac{1}{2}} e^{\lambda(t-s)_{+}} - (-s)_{+}^{H-\frac{1}{2}} e^{\lambda(-s)_{+}}\\
&+\lambda\int_{0}^{t}\left(u-s\right)_{+}^{H-\frac{1}{2}} e^{\lambda(u-s)_{+}}du, \quad s\in\mathbb{R},
\end{split}
\end{equation*}
is called a tempered fractional Brownian motion of the second kind (TFBMI\!I), see \cite{sabzikar2018tempered}.
\end{definition}

The tempered fractional processes TFBMI and TFBMI\!I, defined by \eqref{eq:TFBMdefn} and \eqref{eq:TFBMdefnII}, are obtained by introducing a tempering exponent into the moving average representation of FBM. An alternative tempering approach was introduced in \cite{MolSan}, where the authors modify the autocorrelation function of fractional Gaussian noise (FGN), which is the increment process of FBM. We refer to this process as tempered fractional Brownian motion of the third kind.

\begin{definition}
We consider the overdamped stochastic equation of motion of a particle in a viscous medium
under the influence of a stochastic force $\xi(t)$. A stochastic process \( B_{H,\lambda}^{I\!I\!I} = \{ B_{H,\lambda}^{I\!I\!I}(t) \}_{t \in \mathbb{R}} \) is called a tempered fractional Brownian motion of the third kind (TFBMI\!I\!I) if it satisfies the differential equation:
\begin{equation}\label{eq:TFBMdefnIII}
\frac{dB_{H,\lambda}^{I\!I\!I}(t)}{dt} = \frac{\xi(t)}{m \eta} = \nu(t),
\end{equation}
where $m$ is the particle mass, $\eta$ the friction coefficient and \( \nu(t) \) represents a velocity process with the autocorrelation function given by:
\begin{equation}\label{eq:TFBMdefnIIIcov}
\gamma_H(\tau) = \frac{1}{\Gamma(2H-1)} \tau^{2H-2} e^{-\tau / \tau^*}, \quad \tau > 0,
\end{equation}
where \( \tau^* > 0 \) is a characteristic crossover time scale, and the Hurst parameter satisfies \( \frac{1}{2} \leq H < 1 \). For further details, see \cite{MolSan}.
\end{definition}

The parameter $\tau^*$ is closely related to the application of this process in single-particle experiments in biological cells, as comprehensively described in \cite{MolSan}. Since our aim is to compare all three introduced TFBM types, this parameter will be treated analogously to the tempering parameter used in the two previously discussed processes and will henceforth be referred to as~$\lambda$.

Next, we recall the basic properties of tempered fractional processes \eqref{eq:TFBMdefn}, \eqref{eq:TFBMdefnII}, and \eqref{eq:TFBMdefnIII}. For details, we refer the reader to \cite{meerschaert2013tempered, sabzikar2018tempered, MolSan}.

\begin{proposition}
\begin{itemize}
\item [(1)] TFBMI \eqref{eq:TFBMdefn} and TFBMI\!I \eqref{eq:TFBMdefnII} are Gaussian processes with stationary increments, having the following scaling property:
\begin{equation}\label{eq:Scalling}
	\{X_{H, \lambda}(ct)\}_{t \in \mathbb{R}} \overset{\text{fdd}}{=} \{c^H X_{H, c\lambda}(t)\}_{t \in \mathbb{R}}\text{,}
\end{equation}
where $X_{H,\lambda}$ could be $B^{I}_{H,\lambda}$ or $B^{I\!I}_{H,\lambda}$ (here $\overset{\text{fdd}}{=}$ denotes equality of all finite-dimensional distributions).
\item [(2)] From \eqref{eq:Scalling}
\begin{equation*}
\mathbb{E}\Big[(X_{H,\lambda}(|t|))^2\Big]=|t|^{2H} \mathbb{E}\Big[(X_{H,\lambda|t|}(1))^2\Big]:=|t|^{2H}C^{2}_{t},
\end{equation*}
where the function $C^{2}_{t}$ has the following explicit representation.

(a) If $X_{H,\lambda}=B^{I}_{H,\lambda}$, then
\begin{equation}\label{eq:ct_2_I}
C^{2}_{t}=(C^I_{t})^2 = \mathbb{E}[(B^{I}_{H,\lambda|t|}(1))^2]=\frac{2 \Gamma(2H)}{(2\lambda|t|)^{2H}} -
 \frac{2 \Gamma(H+\frac{1}{2})}{\sqrt{\pi}}\frac{1}{2\lambda|t|^H} K_H(\lambda|t|),
\end{equation}
where $t\neq 0$, and $K_\nu(z)$ is the modified Bessel function of the second kind.

(b) If $X_{H,\lambda}=B^{I\!I}_{H,\lambda}$, then
\begin{equation}\label{eq:ct_2_II}
\begin{split}
C^{2}_{t}&=(C^{I\!I}_{t})^2 = \mathbb{E}[(B^{I\!I}_{H,\lambda|t|}(1))^2]=\frac{(1-2H)\Gamma(H+\frac{1}{2})\Gamma(H)(\lambda t)^{-2H}}{\sqrt{\pi}}\\
&\times \left[1-{}_2F_3 \left(\left\{1, -\frac{1}{2}\right\}, \left\{1-H, \frac{1}{2}, 1\right\}, \frac{\lambda^2t^2}{4}\right) \right]\\
&+\frac{\Gamma(1-H)\Gamma(H+\frac{1}{2})}{\sqrt{\pi}H2^{2H}}
 {}_2F_3\left( \left\{1, H-\frac{1}{2}\right\}, \left\{1, H+1, H+\frac{1}{2}\right\}, \frac{\lambda^2t^2}{4} \right) \text{,}
\end{split}
\end{equation}
where ${}_2F_3$ is the generalised hypergeometric function, that is defined as ${}_pF_q \left(\{a_1, \ldots, a_p\}, \{b_1, \ldots, b_q\}, z\right) = \sum_{n=0}^{\infty} \frac{(a_1)_n \ldots (a_p)_n}{(b_1)_n \ldots (b_q)_n} \frac{z^n}{n!}$ with the use of Pochhammer symbol $((a)_0 = 1\text{; } (a)_n = a(a+1)\ldots(a+n-1)$, $n\geq 1)$ \cite{hypergeom}.

(c) If $X_{H,\lambda}=B^{I\!I\!I}_{H,\lambda}$, then
\begin{equation}
    \label{eq:ct_2_III}
   C^{2}_{t} =(C_{t}^{I\!I\!I})^2 = \mathbb{E}\left[(B^{I\!I\!I}_{H, \lambda}(t))^2\right] = 2 \left(t^{2-2H}E_{1, 3-2H}^{1-2H}\left(-\frac{t}{\tau^*}\right)\right)\text{,}
\end{equation}
where $E_{\alpha, \beta}^{\delta}(z)$ is the three parameter Mittag-Leffler function \cite{MolSan}, defined as
\begin{equation}
    E_{\alpha, \beta}^{\delta}(z) = \sum_{k=0}^{\infty} \frac{(\delta)_k}{\Gamma(\alpha k + \beta} \frac{z^k}{k!}\text{,}
\end{equation}
where $(\delta)_k = \frac{\Gamma(\delta + k)}{\Gamma(\delta)}$ is the Pochhammer symbol.

\item[(3)] Covariance function of the tempered fractional processes have the following form.

(a) TFBMI \eqref{eq:TFBMdefn} with parameter $H$ and $\lambda>0$ has the covariance function
\begin{equation}
\label{eq:tfbm1_cov}
	\Cov\left[B^I_{H, \lambda}(t), B^I_{H, \lambda}(s)\right] = \frac{1}{2}\Big[C^{2}_t |t|^{2H} + C^{2}_s |s|^{2H} - C^{2}_{t-s}|t-s|^{2H}\Big]
\end{equation}
for any $s,t\in\mathbb{R}$ where $C^{2}_{t}=(C^{I}_{t})^{2}$ is given by \eqref{eq:ct_2_I}.

(b) TFBMI\!I \eqref{eq:TFBMdefnII} with parameter $H$ and $\lambda>0$ has the covariance function
\begin{equation}
\label{eq:tfbm2_cov}
	\Cov\left[B^{I\!I}_{H, \lambda}(t), B^{I\!I}_{H, \lambda}(s)\right] = \frac{1}{2}\Big[C^{2}_t |t|^{2H} + C^{2}_s |s|^{2H} - C^{2}_{t-s}|t-s|^{2H}\Big]
\end{equation}
for any $s,t\in\mathbb{R}$ where $C^{2}_{t}=(C^{I\!I}_{t})^{2}$ is given by \eqref{eq:ct_2_II}.

(c) TFBMI\!I\!I \eqref{eq:TFBMdefnIII} with parameter $H>\frac{1}{2}$ and $\lambda>0$ has the covariance function
\begin{equation}
\label{eq:tfbm3_cov}
	\Cov\left[B^{I\!I\!I}_{H, \lambda}(t), B^{I\!I\!I}_{H, \lambda}(s)\right] = \frac{1}{2}\Big[C^{2}_t  + C^{2}_s  - C^{2}_{t-s}\Big]
\end{equation}
for any $s,t\in\mathbb{R}$ where $C^{2}_{t}=(C^{I\!I\!I}_{t})^{2}$ is given by \eqref{eq:ct_2_III}.
\end{itemize}
\end{proposition}

Next, we recall the definitions of tempered fractional Gaussian noise (TFGN) of the first kind (TFGNI), TFGN of the second kind (TFGNI!I), and
TFGN of the third kind (TFGNI\!I\!I).

For simplicity, denote $\alpha=H-\frac{1}{2}$. Given a TFBMI \eqref{eq:TFBMdefn}, we define TFGNI:
\begin{equation*}\label{eq:TFGNdefn}
\beta^{I}_{\alpha,\lambda}(j)=B^{I}_{H,\lambda}(j+1)-B^{I}_{H,\lambda}(j)\quad\text{for}\quad  j\in \mathbb{Z}.
\end{equation*}
It follows easily from \eqref{eq:TFBMdefn} that TFGNI has the moving average representation:
\begin{equation}\label{eq:TFGNImoving}
\begin{split}
\beta^{I}_{\alpha,\lambda}(j)&= \int_{\mathbb{R}}g^{I}_{\lambda,\alpha,j}(x)B(dx)\\
&={\int_{\mathbb{R}}\left[e^{-\lambda(j+1-x)_{+}}(j+1-x)_{+}^{\alpha}-e^{-\lambda(j-x)_{+}}(j-x)_{+}^{\alpha} \right]B(dx)}.
\end{split}
\end{equation}
Along the same lines, TFGNI\!I can be defined as follows:
\begin{equation*}
\beta^{I\!I}_{\alpha,\lambda}(j)=B^{I\!I}_{H,\lambda}(j+1)-B^{I\!I}_{H,\lambda}(j)\quad\text{for  $j\in \mathbb{Z}$.}
\end{equation*}
It follows from  \eqref{eq:TFBMdefnII} that  a TFGNI\!I  has the moving average representation
\begin{equation}\label{eq:TFGNIImoving}
\begin{split}
\beta^{I\!I}_{\alpha,\lambda}(j)&= \int_{\mathbb{R}}g^{I\!I}_{\lambda,\alpha,j}(x)B(dx)=\int_\rr \Big[ e^{-\lambda(j+1-x)_{+}}(j+1-x)_{+}^{\alpha}-e^{-\lambda(j-x)_{+}}(j-x)_{+}^{\alpha}\\
&\qquad\qquad\qquad\quad+\lambda \int_{j}^{j+1} e^{-\lambda(s-x)_{+}}(s-x)_{+}^{\alpha} ds\Big]  B(dx).
\end{split}
\end{equation}

So, let $\beta^{I}_{\alpha,\lambda}(j)$ and $\beta^{I\!I}_{\alpha,\lambda}(j)$  be the stationary sequences given by \eqref{eq:TFGNImoving} and \eqref{eq:TFGNIImoving} respectively. Denote
\begin{equation}\label{eq:covar}
\begin{split}
\gamma^J (k):&=\mathbb{E}[\beta^{J}_{\alpha,\lambda}(0) \beta^{J}_{\alpha,\lambda}(k)]\\
&=\abs{k+1}^{2H}(C^{J}_{\vert k \vert +1})^2-2\abs{k}^{2H} (C^{J}_{\vert k\vert})^2+\abs{k-1}^{2H}(C^{J}_{\vert k-1 \vert })^2,\; J=I,I\!I,
\end{split}
\end{equation}
where the normalizing constants $C^J_t$ are presented in \eqref{eq:ct_2_I} and \eqref{eq:ct_2_II}.

We also define TFGNI\!I\!I:
\begin{equation*}\label{eq:TFGNIIIdefn}
\beta^{I\!I\!I}_{\alpha,\lambda}(j)=B^{I\!I\!I}_{H,\lambda}(j+1)-B^{I\!I\!I}_{H,\lambda}(j)\quad\text{for}\quad  j\in \mathbb{Z}.
\end{equation*}
with the covariance function
\begin{equation}
	\label{eq:incr_acvf3}
	\gamma_{I\!I\!I}(j) = \frac{1}{2} \left\{(C^{I\!I\!I}_{j+1})^2 - 2(C^{I\!I\!I}_j)^2 + (C^{I\!I\!I}_{j-1})^2\right\}\text{,}
\end{equation}
where $C_t^{I\!I\!I}$ is presented in $\eqref{eq:ct_2_III}$.

To demonstrate the dynamics of the analysed TFBMs, let us now present so-called quantile lines calculated for realisations of those processes. The idea of the quantile lines is as follows.
Let us assume that we observe $M$ samples of length $N$ and denote their values by $\{Z_n^k\}$, $n=1,2,\ldots N$, $k=1,...,M$,  and $0<p_j<1$, $j=1,...,J$ are given probabilities.  It is possible to derive  estimators of the corresponding quantiles $q_j(n)= F_n^{-1}(p_j)$, where $F_n=F_n(x)$ denotes CDF of the random variable $Z_n$ represented by the statistical  sample $Z_n^k$, $k=1,...,M$. In this way we obtain the approximation of the quantile lines, i.e., the curves
$$q_j=q_j(n)$$
 defined by the condition
$$
P\{Z_n\leq q_j(n)\}=p_j.
$$
In layman terms, the quantile lines represent the value $q_j$ for which $p_j*100\%$ of the data are below at a certain time point $n$. For a stationary process the quantile lines $q_j(n) =$ const, whereas for a $H$-self-similar process they behave like $n^H$.

\begin{figure} [!h]
    \centering
    \includegraphics[width=0.99\linewidth]{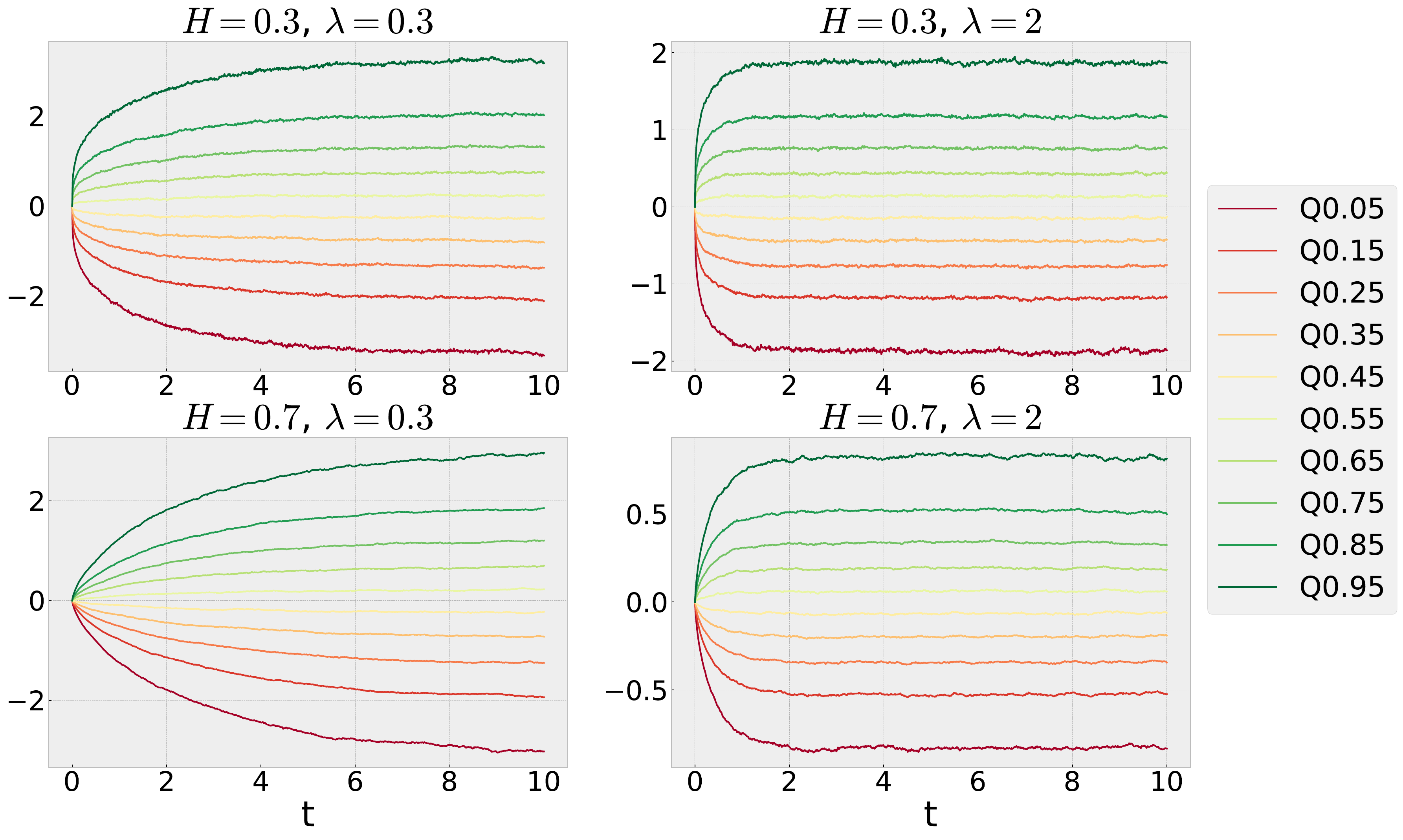}
    \caption{Quantile lines for the simulated $N=1000$ trajectories of TFBMI for $T = 10$.}
    \label{fig:quantile_lines_tfbm1}
\end{figure}

\begin{figure} [!h]
    \centering
    \includegraphics[width=0.99\linewidth]{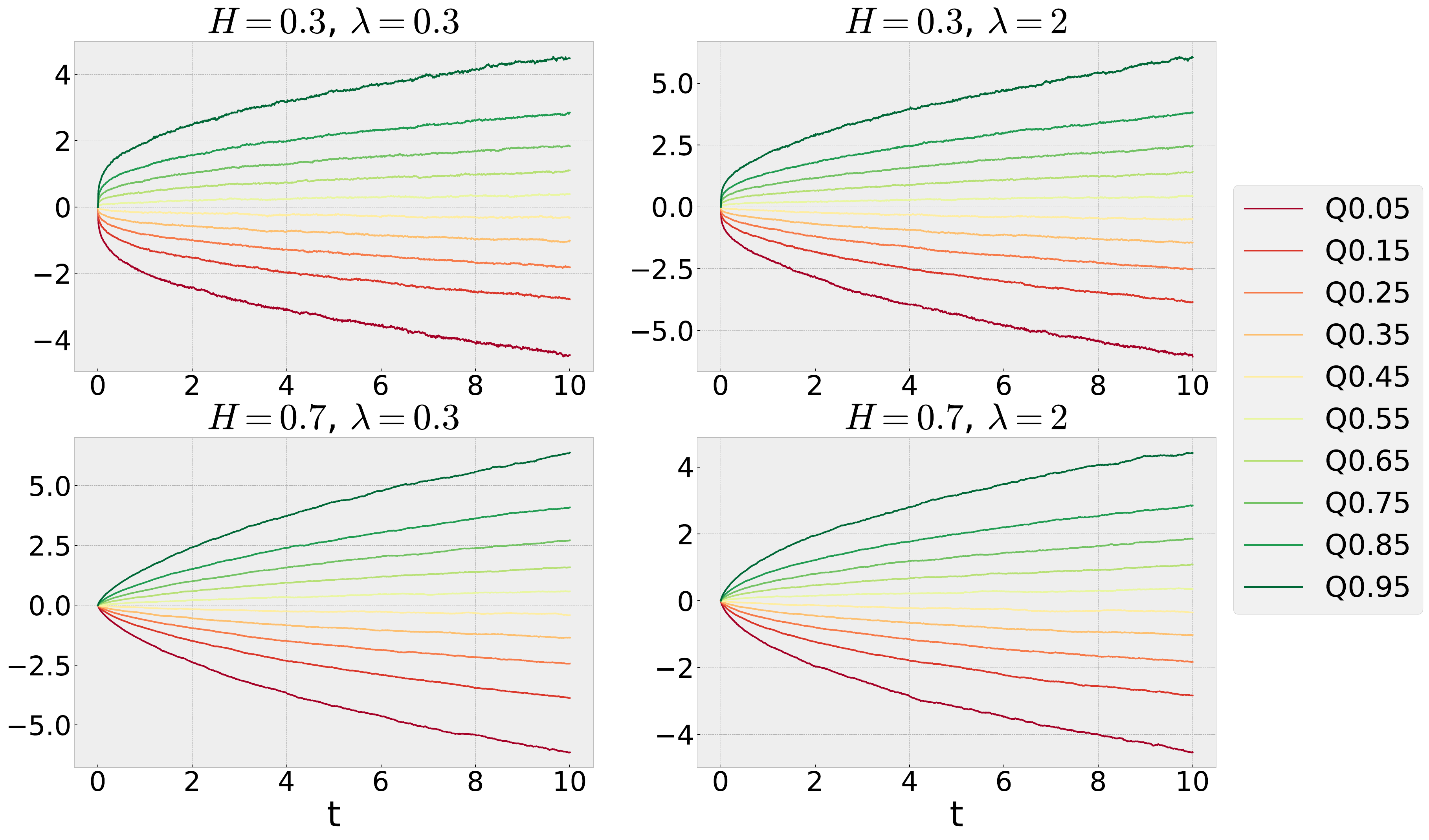}
    \caption{Quantile lines for the simulated $N=1000$ trajectories of TFBMI\!I for $T = 10$.}
    \label{fig:quantile_lines_tfbm2}
\end{figure}

\begin{figure} [!h]
    \centering
    \includegraphics[width=0.99\linewidth]{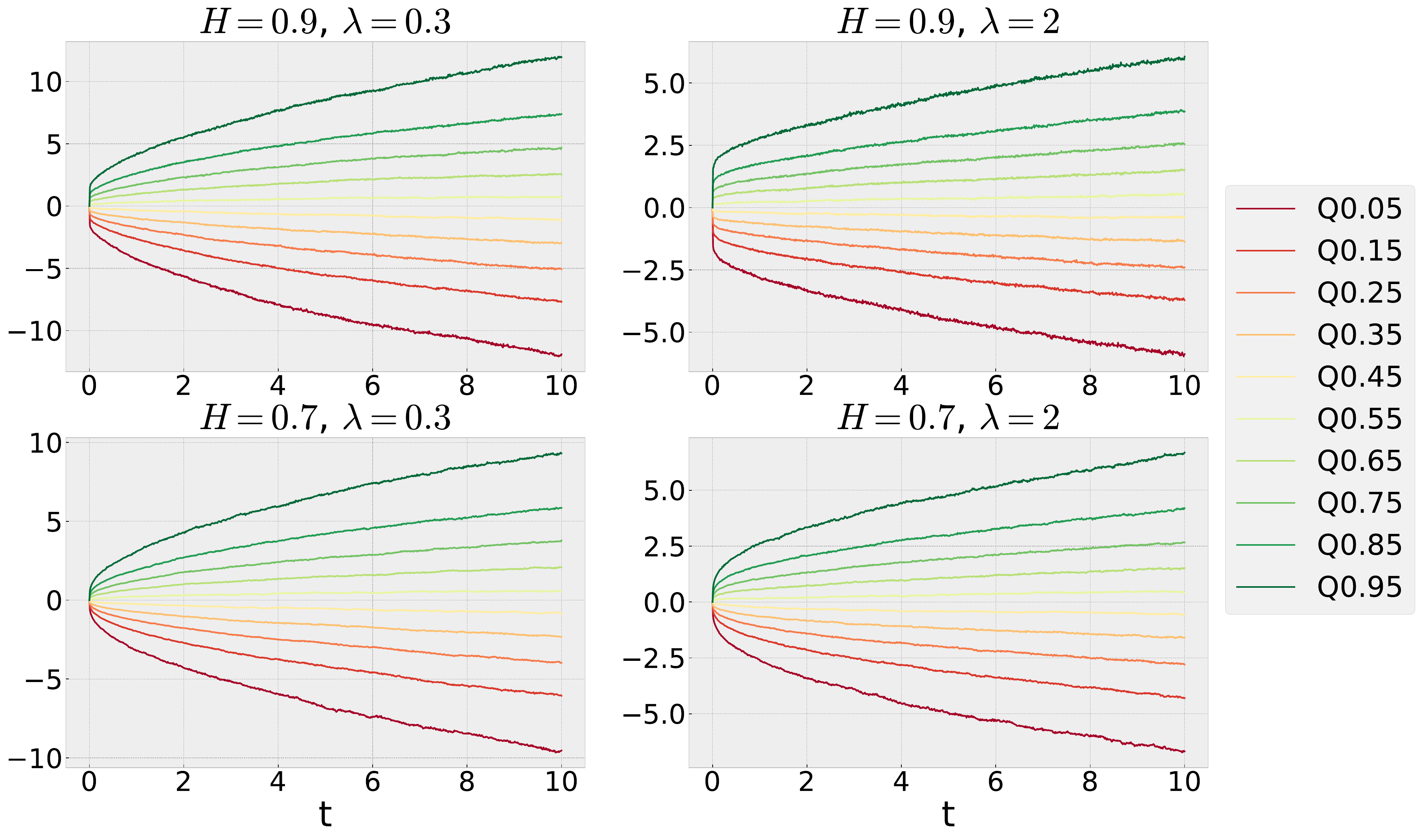}
    \caption{Quantile lines for the simulated $N=1000$ trajectories of TFBMI\!I\!I for $T = 10$.}
    \label{fig:quantile_lines_tfbm3}
\end{figure}

In Figures \ref{fig:quantile_lines_tfbm1}, \ref{fig:quantile_lines_tfbm2} and \ref{fig:quantile_lines_tfbm3} we depict quantile lines
for TFBMI, FBMI\!I and TFBMI\!I\!I, respectively. The parameter $H \in \{0.3, 0.7\}$ for TFBMI and TFBMI\!I, $H \in \{0.7, 0.9\}$ for TFBMI\!I\!I, and $\lambda \in \{0.3, 2\}$. We chose $N=M=1000$ and the time horizon $T=10$ for simulation purposes. We can observe that the behaviour of TFBMI is very different from other processes, namely its quantile lines become flat. This is due to the fact that TFBMI is asymptotically stationary; see \cite{MolSan}.
We note that the trajectories were generated using the Cholesky or Davies--Harte method depending on a specific case, as each of the methods has some limitations \cite{DaviesHarte,Dieker}. The Davies-Harte approach was used more frequently due to computational complexity advantage; however, when we failed to meet the assumptions of this method, the Cholesky method was utilised.

\section{Tempered processes testing based on quadratic form statistics}
\label{chap:test_intro}
\vskip 0.3cm
The testing methodology for Gaussian processes, introduced in~\cite{Chaos}, can be used with different statistics, which have a quadratic form representation.
We consider a random sample of length $N$ from a centered Gaussian process $\{X(t),$ $t \geq 0\}$:
\begin{equation}
	\mathbb{X}_N = \{X(1), X(2), \ldots, X(N)\}\text{,}
\end{equation}
with the covariance matrix of $\mathbb{X}_N$ given by $\Sigma_N$.
The idea of the methodology is based on the choice of a statistic that can be represented in a quadratic form:
\begin{equation}
	\label{eq:quadratic_form}
	S_N(\tau) = \mathbb{X}_N\mathbb{A}(\tau) \mathbb{X}_N^T\text{,}
\end{equation}
where $\mathbb{A}(\tau) = \left[a_{i, j}\right]_{i=1, \ldots, N, j=1, \ldots, N}$ is some matrix, that depends on the time lag $\tau$, where $\tau \in \{0, 1, \ldots, N-1\}$ \cite{Chaos}. 

The quadratic form has a generalised $\chi^2$ distribution:
\begin{equation}
\label{eq:test_stat}
	S_N(\tau) \overset{\mathrm{d}}{=} \sum_{i=1}^{N-\tau} \lambda_i(\tau) U_i\text{,}
\end{equation}
where
\begin{itemize}
	\item $U_i$'s are independent identically distributed (i.i.d.) random variables that have the generalised $\chi^2$ distribution with one degree of freedom,
	\item $\lambda_i(\tau)$ are the eigenvalues of the matrix $\Sigma_N^{1/2}\mathbb{A}(\tau)\Sigma_N^{1/2}$.
\end{itemize}

Following \cite{Chaos}, let us now present three statistics used in this paper.

\subsection{Autocovariance function}
\label{sec:test_acvf}
\vskip 0.3cm
We start with the sample autocovariance function (ACVF) statistic that has the form:
\begin{equation}
	\label{eq:acvf}
	ACVF(\tau) = \frac{1}{N-\tau} \sum_{i=1}^{N-\tau} X(i+\tau)X(i)\text{.}
\end{equation}

In this case the matrix $\mathbb{A}(\tau)$ is either a diagonal matrix (for $\tau=0$) or the Toeplitz matrix with only two non-zero subdiagonals. It can be presented as:
\begin{equation}
\label{eq:a_k_matrix_acvf1}
	a_{i, j} =
	\begin{cases}
		\frac{1}{2}\frac{1}{N-\tau} & \text{ for every } i \text{, } j \text{ such that } |i-j| = \tau \text{,} \\
		0 & \text{ otherwise,}
	\end{cases}
\end{equation}
for $\tau \in \{1, 2, \ldots, N-1\}$, and
\begin{equation}
\label{eq:a_k_matrix_acvf2}
	a_{i, j} =
	\begin{cases}
		\frac{1}{N} & \text{ for every } i \text{, } j \text{ such that } i = j \text{,} \\
		0 & \text{ otherwise,}
	\end{cases}
\end{equation}
for $\tau=0$ \cite{Chaos}.
Finally, we note that as the increment process of TFBMI, so TFGNI, is also a centred Gaussian process. We will use it for the $ACVF$ based test since we found that this approach is more stable than the application of the test to the original process.


\subsection{Detrended moving average}
\label{sec:test_dma}
\vskip 0.3cm
The detrended moving average (DMA) statistic introduced in \cite{allesio_2002} is another statistic that can be represented in a quadratic form. It is defined as
\begin{equation}
\label{eq:dma}
	DMA(\tau) = \frac{1}{N-\tau} \sum_{i=1}^{N} \left(X(i) - \tilde{X}_{\tau}(i)\right)^2\text{,}
\end{equation}
where
\begin{equation}
	\tilde{X}_{\tau}(i) = \frac{1}{\tau} \sum_{j=0}^{\tau-1} X(i-j)
\end{equation}
is the moving average of $\tau$ observations: $X(i), X(i+1), \ldots, X(i-\tau+1)$.

In \cite{sikora_chaos_2018} it was proven that the DMA statistic has a quadratic form representation (\ref{eq:quadratic_form}) but not in terms of the original trajectory $\mathbb{X}_N$ but in terms of the detrended process $Y(i):=X(i+\tau-1)-\bar{X}^{\tau}(i+\tau-1),$ for $i=1,2,\ldots,N-\tau+1$, namely
\begin{equation}\label{dma}
DMA(\tau)=\frac{1}{N-\tau+1}\mathbb{Y}_{N-\tau+1}\mathbb{Y}_{N-\tau+1}^T,
\end{equation}
where $\mathbb{Y}_{N}(\tau)=\{Y(1),Y(2),\ldots,Y(N-\tau+1)\}.$

The trajectory $\mathbb{Y}_N(\tau)$ is still a centred Gaussian process with the covariance matrix explainable in terms of covariance of the original sample trajectory:
\begin{eqnarray}
\label{eq:sigma_tilde_dma}
\tilde{\Sigma}_N=\mathbb{E}[Y(j)Y(k)]&=&\left(1-\frac{1}{\tau}\right)^2 \mathbb{E}[X(j+N-1)X(k+N-1)]\nonumber\\
&+&\left(\frac{1}{\tau^2}-\frac{1}{\tau}\right)\left[\sum_{m=k}^{k+\tau-2}\mathbb{E}[X(j+\tau-1)X(m)]
\right.\nonumber\\
&+& \left. \sum_{l=j}^{j+n-2}\mathbb{E}[X(l)X(k+\tau-1)\right]\nonumber\\&+&\frac{1}{\tau^2}\sum_{j\leq l\leq j+\tau-2}\sum_{k\leq m\leq k+\tau-2}\mathbb{E}[X(l)X(m)].
\end{eqnarray}

As a consequence, we represent DMA statistic in a quadratic form:
\begin{equation}
	DMA(\tau) = \frac{1}{N - \tau +1} \sum_{i=\tau}^{N} Y^2(i-\tau+1) = \frac{1}{N - \tau +1} \mathbb{Y}_{N-\tau+1} \mathbb{Y}^T_{N-\tau+1}\text{,}
\end{equation}
where $\mathbb{Y}^T_{N-\tau+1}$ is a transpose of $\mathbb{Y}_{N-\tau+1}$ and is a vertical vector.

Therefore, we can represent it as:
\begin{equation}
	S_N(\tau) = \frac{1}{N - \tau +1} \sum_{i=1}^{N-\tau +1} \lambda_i(\tau)U_i\text{,}
\end{equation}
where $\lambda_i(\tau)$s are the eigenvalues of the covariance matrix $\tilde{\Sigma}_N$, $U_i$s are the random variables from a generalized $\chi^2$ distribution with one degree of freedom and the matrix $\mathbb{A}(\tau)$ in Eq. (\ref{eq:quadratic_form}) is an identity matrix \cite{sikora_chaos_2018}.


\subsection{Time average mean-squared displacement}
\label{sec:test_tamsd}
\vskip 0.3cm
The last considered statistic is the time averaged mean-squared displacement (TAMSD) which is defined as
\begin{equation}
\label{eq:tamsd}
	TAMSD(\tau) = \frac{1}{N-\tau} \sum_{i=1}^{N-\tau} \left(X(i+\tau) - X(i)\right)^2\text{,}
\end{equation}
for time lag $\tau \in \{1, 2, \ldots, N-1\}$.

TAMSD is another statistic that can be represented in quadratic form given in Eq. $\eqref{eq:quadratic_form}$ \cite{sikoraburnecki2017}. The matrix $\mathbb{A}(\tau)$ is once again a specific matrix that depends on the time lag $\tau$, which, this time, is defined as $\mathbb{A}(\tau) = \{a_{i, j}\}_{i, j=1}^N$, where
\begin{equation}
	\label{eq:a_tamsd}
	a_{i, j} = \frac{2}{N-\tau} \left(m_i\mathbb{I}(i=j) - \mathbb{I}(|i-j|=\tau)\right)\text{.}
\end{equation}
The $m_i$'s in $\eqref{eq:a_tamsd}$ are defined for $\tau \leq \frac{N}{2}$ as
\begin{equation}
	m_i =
	\begin{cases}
		1, & \text{ for } i \leq \tau \text{ or } i \geq N-\tau, \\
		2, & \text{ for } \tau < i < N-\tau\text{,}
		
	\end{cases}
\end{equation}
and for $\tau > \frac{N}{2}$ as
\begin{equation}
	m_i =
	\begin{cases}
		1, & \text{ for } i \geq \tau \text{ or } i \leq N-\tau, \\
		0, & \text{ for } \tau < i < N-\tau\text{.}
	\end{cases}
\end{equation}


\subsection{Testing procedure for tempered fractional Brownian motions}
\label{sec:test_algorithm}
\vskip 0.3 cm

Let us now focus on TFBMs. We consider an empirical sample $\mathbb{B}_{H, \lambda} = \{B_{H, \lambda}^J(1), B_{H, \lambda}^J(2), \ldots, B_{H, \lambda}^J(N)\}$ and a two-sided test with the following hypotheses:
\begin{itemize}
	\item the null hypothesis: $\mathcal{H}_0$: $\mathbb{B}_{H, \lambda}^J$ is a trajectory of TFBM of type $J$ with parameters $H$ and $\lambda$,
	\item the alternative hypothesis: $\mathcal{H}_1$: $\mathbb{B}_{H, \lambda}^J$ is not a trajectory of TFBM of type $J$ with parameters $H$ and $\lambda$,
\end{itemize}
where $J = I, I\!I, I\!I\!I$, so we consider all three types of TFBM.

The null hypothesis is rejected when the test statistic is extreme. For a given significance level $c$ we can present the acceptance region as:
\begin{equation}
\label{eq:acceptance_region}
	\left[Q_{c/2}(N, \tau), Q_{1-c/2}(N, \tau)\right]\text{,}
\end{equation}
where $Q_{p}(N, \tau)$ is the quantile of order $p$ of the generalised $\chi^2$ distribution defined by $\eqref{eq:test_stat}$. The acceptance region also allows us to calculate the power of the proposed test.


Let us now present Algorithm \ref{alg:test_stat} for the generation of test statistics for the three statistics and Algorithm \ref{alg:test_power}
 for the calculation of the power of the test that will be used later to assess the performance of the tests.

\begin{algorithm}
\begin{algorithmic}[1]
\State Fix test parameters:
\begin{itemize}
    \item $H$ -- Hurst index corresponding to $\mathcal{H}_0$,
    \item $\lambda$ -- tempering parameter corresponding to $\mathcal{H}_0$,
    \item $\tau$ -- time lag.
\end{itemize}
\State Calculate $\lambda_i(\tau)$ for $i = 1, \ldots, N-1$ that are the eigenvalues of a matrix $\mathbb{B}_N(\tau)$ that is specific for each statistic:
\begin{itemize}
    \item $\textbf{ACVF}$: $\mathbb{B}_N(\tau) = \Sigma_N^{1/2}\mathbb{A}(\tau)\Sigma_N^{1/2}$ with the coviariance matrix $\Sigma_N$ of the corresponding TFGN and $\mathbb{A}(\tau)$ described by $\eqref{eq:a_k_matrix_acvf1}$ and $\eqref{eq:a_k_matrix_acvf2}$,
    \item $\textbf{DMA}$: $\mathbb{B}_N(\tau) = \tilde{\Sigma}_N$, with $\tilde{\Sigma}_N$ being presented in $\eqref{eq:sigma_tilde_dma}$.
    \item $\textbf{TAMSD}$: $\mathbb{B}_N(\tau) = \Sigma_N^{1/2}\mathbb{A}(\tau)\Sigma_N^{1/2}$ with $\Sigma_N$ being the covariance matrix of TFBM and $\mathbb{A}(\tau)$ described by $\eqref{eq:a_tamsd}$.
\end{itemize}
\State Generate $L$ times a sample $\mathbf{U}^l$, $l = 1, 2, \ldots, L$ of i.i.d. random variables from the generalized $\chi^2$ distribution with one degree of freedom.
\State Calculate the test statistic as:
\begin{equation}
    S_N(\tau) = \sum_{i=1}^{N-\tau} \lambda_i(\tau)U_i\text{.}
\end{equation}
\caption{Test statistic generation}
\label{alg:test_stat}
\end{algorithmic}
\end{algorithm}



\begin{algorithm}
    \begin{algorithmic}[1]
    \State Simulate a trajectory of the tested process (in most cases that would be alternative process, but sometimes we test against the same process but different parameters) and calculate appropriate statistic denoted by $\eta$.
    \State Calculate the test statistic using $L$ Monte Carlo simulations and Algorithm \ref{alg:test_stat}.
    \State Calculate appropriate quantiles for a given significance level c as shown in $\eqref{eq:acceptance_region}$. If $\eta \in [Q_{c/2}(N, \tau), Q_{1-c/2}(N, \tau)]$ then the counting variable is set to $0$ otherwise to 1.
    \State The power is approximated by repeating former steps $M$ times and calculating mean of counting variables.
    \caption{Test power calculation}
    \label{alg:test_power}
    \end{algorithmic}
\end{algorithm}



\section{Power simulation study}
\label{sec:power_results}
\vskip 0.3cm

With Algorithms \ref{alg:test_stat} and \ref{alg:test_power} provided, we can now move on to the test power study to examine the test performance for different scenarios. Using $M = 10\;000$ Monte Carlo simulations, we create the plots of the test power.
We present the results for each of the TFBMs separately, that is, for each kind of TFBM and a fixed set of parameters $(H_0, \lambda_0)$, we simulate the process and two other TFBMs and FBM with various parameters $H$ and calculate the power of the test,  see Figures \ref{fig:tfbm1_power_var_H_H1_l1}, \ref{fig:tfbm1_power_var_H_H2_l1}, \ref{fig:tfbm2_power_var_H_H1_l1}, \ref{fig:tfbm2_power_var_H_H2_l1}, \ref{fig:tfbm3_power_var_H_H1_l1} and \ref{fig:tfbm3_power_var_H_H2_l1}.

We also present the test results for each of the TFBMs, fixed set of parameters $(H_0, \lambda_0)$ and  the test alternatives being the same TFBM model but with different $\lambda$ values, see Figures \ref{fig:tfbm1_power_var_lambd}, \ref{fig:tfbm2_power_var_l} and \ref{fig:tfbm3_power_var_l_H1}.

For TFBMI and TFBMI\!I we examine $H_0 \in \{0.3, 0.7\}$ and $\lambda_0 \in \{0.3, 2\}$, whereas for TFBMI\!I\!I, due to its constraints, we choose $H_0 \in \{0.7, 0.9\}$ and $\lambda_0 \in \{0.3, 2\}$.
For each case, we take into account three different statistics: ACF, DMA, and TAMSD, and two different sample lengths $N \in \{200, 1000\}$.

For the analysis,  we also selected specific values of the parameter $\tau$: $\tau = 1$,  $\tau = 2$ and $\tau = 1$ for ACF, DMA and TAMSD based tests, respectively.
Those values were chosen on the basis of our preliminary studies, which showed that such a choice leads to the highest power values.


\subsection{TFBMI}
\label{sec:tfbm1_power_results}
\vskip 0.3cm

We start the test power study with the null hypothesis $\mathcal{H}_0$: the sample is a trajectory of TFBMI with parameters $H_0$ and $\lambda_0$.
\begin{figure} [!h]
    \centering
    \includegraphics[width=0.99\linewidth]{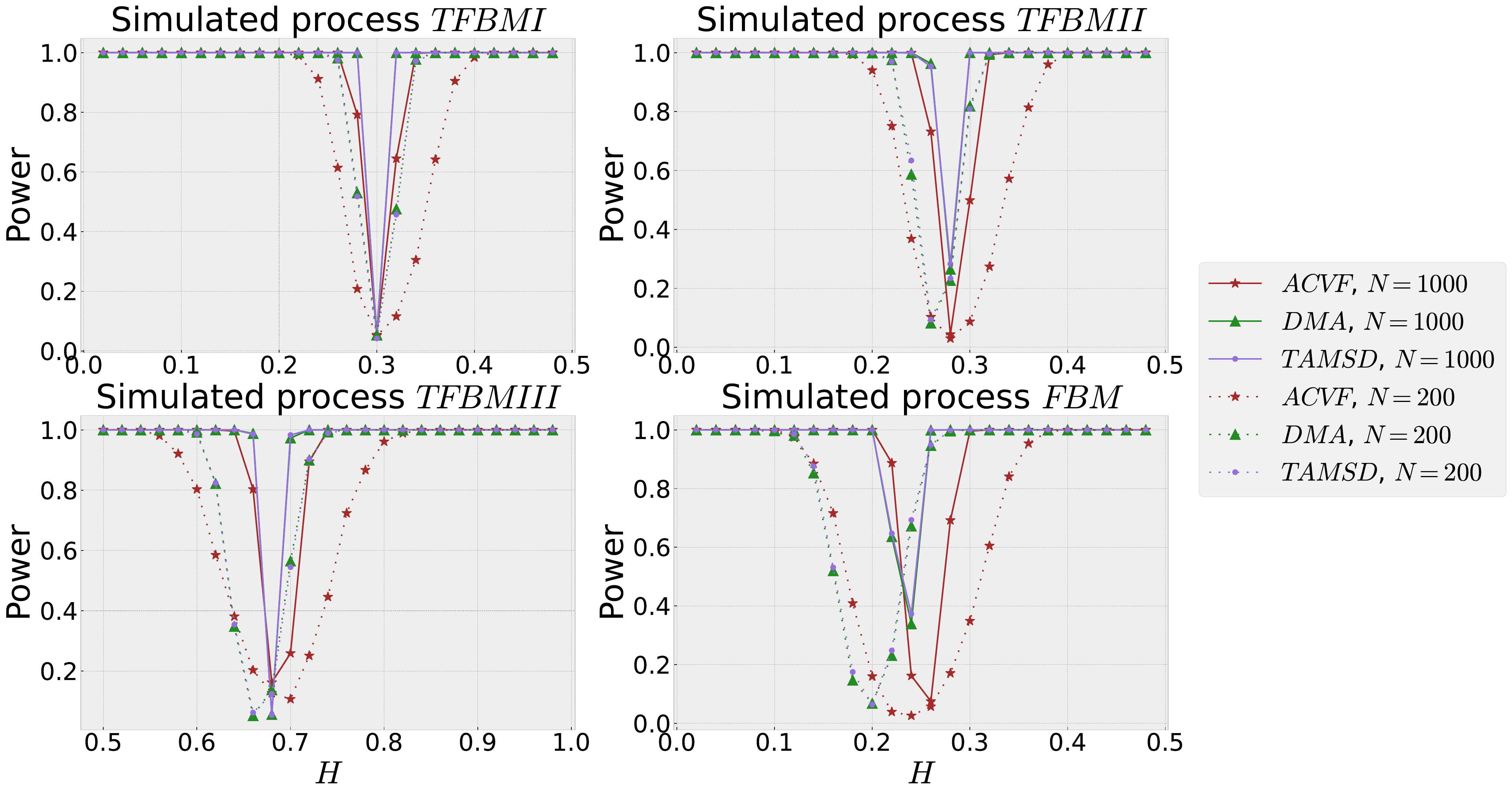}
    \caption{TFBMI test power for $H_0 = 0.3$ and $\lambda_0 = 0.3$ calculated for TFBMI, TFBMI\!I, TFBMI\!I\!I and FBM trajectories with different $H$ values for two sample lengths ($N \in \{200, 1000\}$).}
    \label{fig:tfbm1_power_var_H_H1_l1}
\end{figure}
\begin{figure} [!h]
    \centering
    \includegraphics[width=0.99\linewidth]{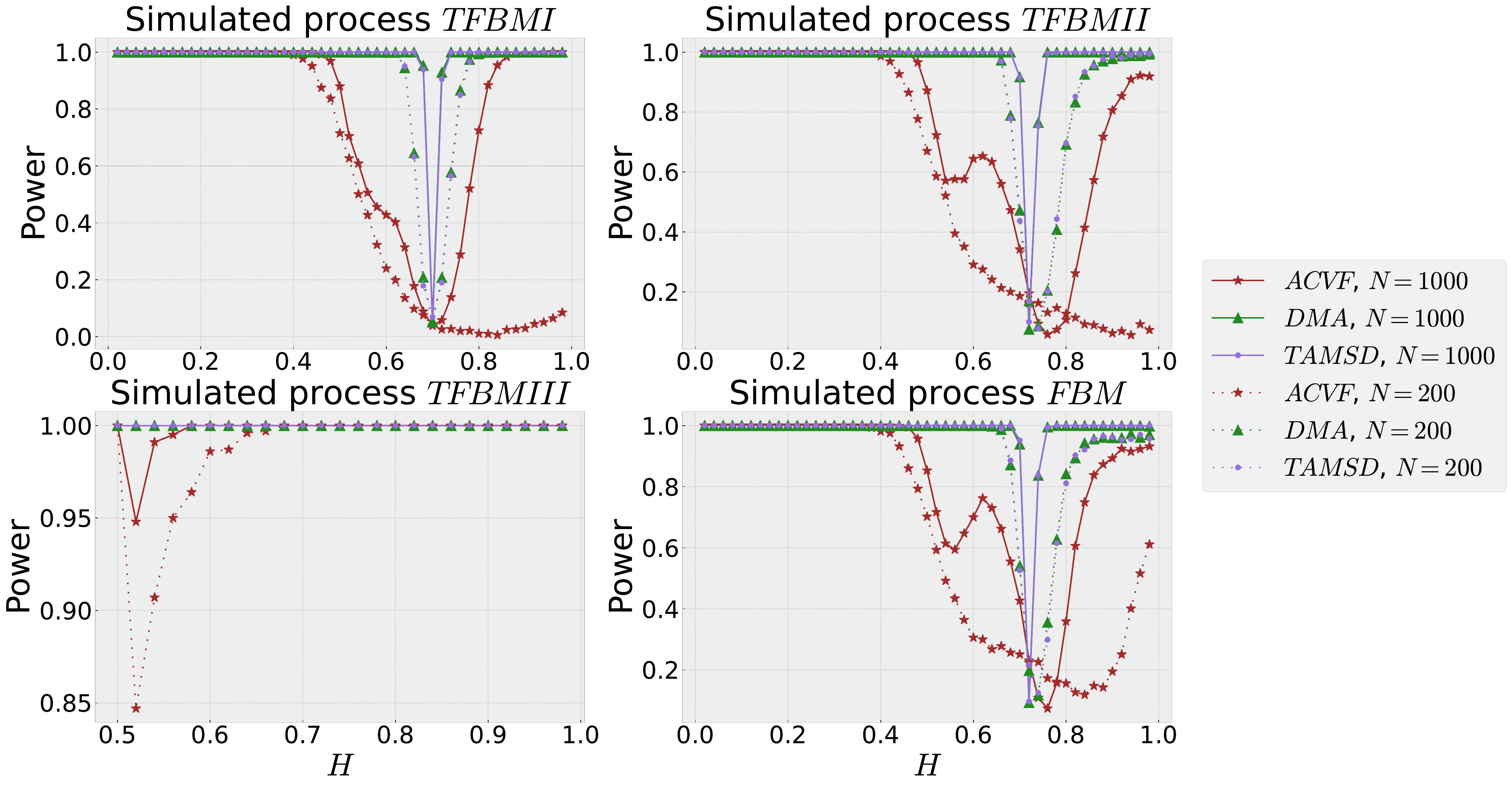}
    \caption{TFBMI test power for $H_0 = 0.7$ and $\lambda_0 = 0.3$ generated for TFBMI, TFBMI\!I, TFBMI\!I\!I and $FBM$ trajectories with different $H$ values for two sample lengths ($N \in \{200, 1000\}$).}
    \label{fig:tfbm1_power_var_H_H2_l1}
\end{figure}
We can observe in Figures~\ref{fig:tfbm1_power_var_H_H1_l1} and \ref{fig:tfbm1_power_var_H_H2_l1} that the ACF based test most often provides the worst results. Other statistics give similar results. Moreover, we are also able to clearly notice the difference in the test performance for the two examined sample lengths: the larger $N$ leads to a much higher power.


Moreover, in the left top panel of Figure \ref{fig:tfbm1_power_var_H_H1_l1} we can see that the power of the tests quickly increases to 1 when $H$ of TFMI departs from the true value $H_0=0.3$.
When the alternatives are different processes (other panels in Figure \ref{fig:tfbm1_power_var_H_H1_l1}), we
can observe a similar behaviour, only for TFBMI\!I and FBM the lowest power (equal to the significance level of the test) is reached for $H$ slightly less than $0.3$ and for TFBMI\!I\!I the lowest value is for $H$ slightly less than $0.7$.  The latter phenomenon is due to the fact that TFBMI\!I\!I is completely different process from other TFBMs (e.g., it is well-defined only for $H>0.5$).

In Figure \ref{fig:tfbm1_power_var_H_H2_l1}, which corresponds to $H=0.7$, the situation differs especially under the alternative of TFBMI\!I\!I  (left bottom panel), where the false null hypothesis is always rejected with high probability.  Moreover, now the power of the test is the lowest under the alternatives of TFBMI\!I and FBM for $H$ slightly greater than $0.7$.

We also performed the same analyses for $\lambda_0 = 2$, but the results were very similar, so we decided not to include them in the paper.

\begin{figure} [!h]
    \centering
    \includegraphics[width=0.99\linewidth]{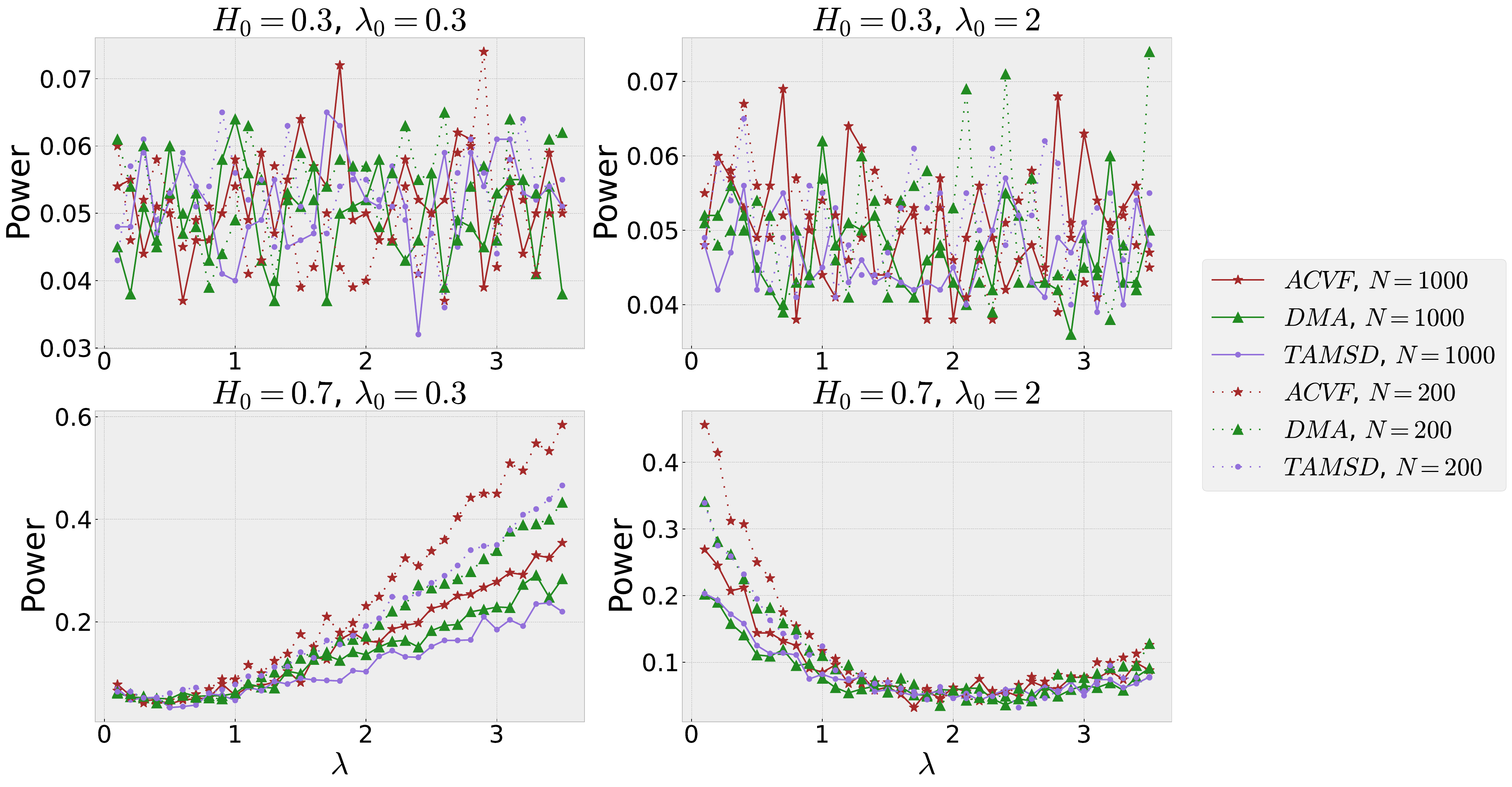}
    \caption{TFBMI test power for $H_0 \in \{0.3, 0.7\}$ and $\lambda_0 \in \{0.3, 2\}$ calculated for TFBMI trajectories with different $\lambda$ values for two sample lengths ($N \in \{200, 1000\}$).}
    \label{fig:tfbm1_power_var_lambd}
\end{figure}

Finally, we also examine the power of the test for fixed $H$ and varying tempering parameter $\lambda$, see Figure $\ref{fig:tfbm1_power_var_lambd}$. This time we examine it only for the samples from the same TFBM type, so in this case TFBMI.
Figure $\ref{fig:tfbm1_power_var_lambd}$ shows a major difference in test power between two examined Hurst index values. For $H_0: H = 0.3$ the test power oscillates around the significance value $0.05$. It shows that the presented testing methodology will most likely result in incorrect acceptance of the null hypothesis when $\lambda \neq \lambda_0$. Also, for the first time we observe that the sample length $N$ does not appear to have an impact on the results. In contrast, for $H_0: H = 0.7$ the power is more dependent on $\lambda$ values. When it is close to $\lambda_0$ the power is the lowest. For other $\lambda$'s the power reaches 0.6. As a result, in this case, we are more likely to reject the false null hypothesis, when only the value $\lambda$ is incorrect.

\subsection{TFBMI\!I}
\label{sec:tfbm2_power_results}
\vskip 0.3cm


\begin{figure} [!h]
    \centering
    \includegraphics[width=0.99\linewidth]{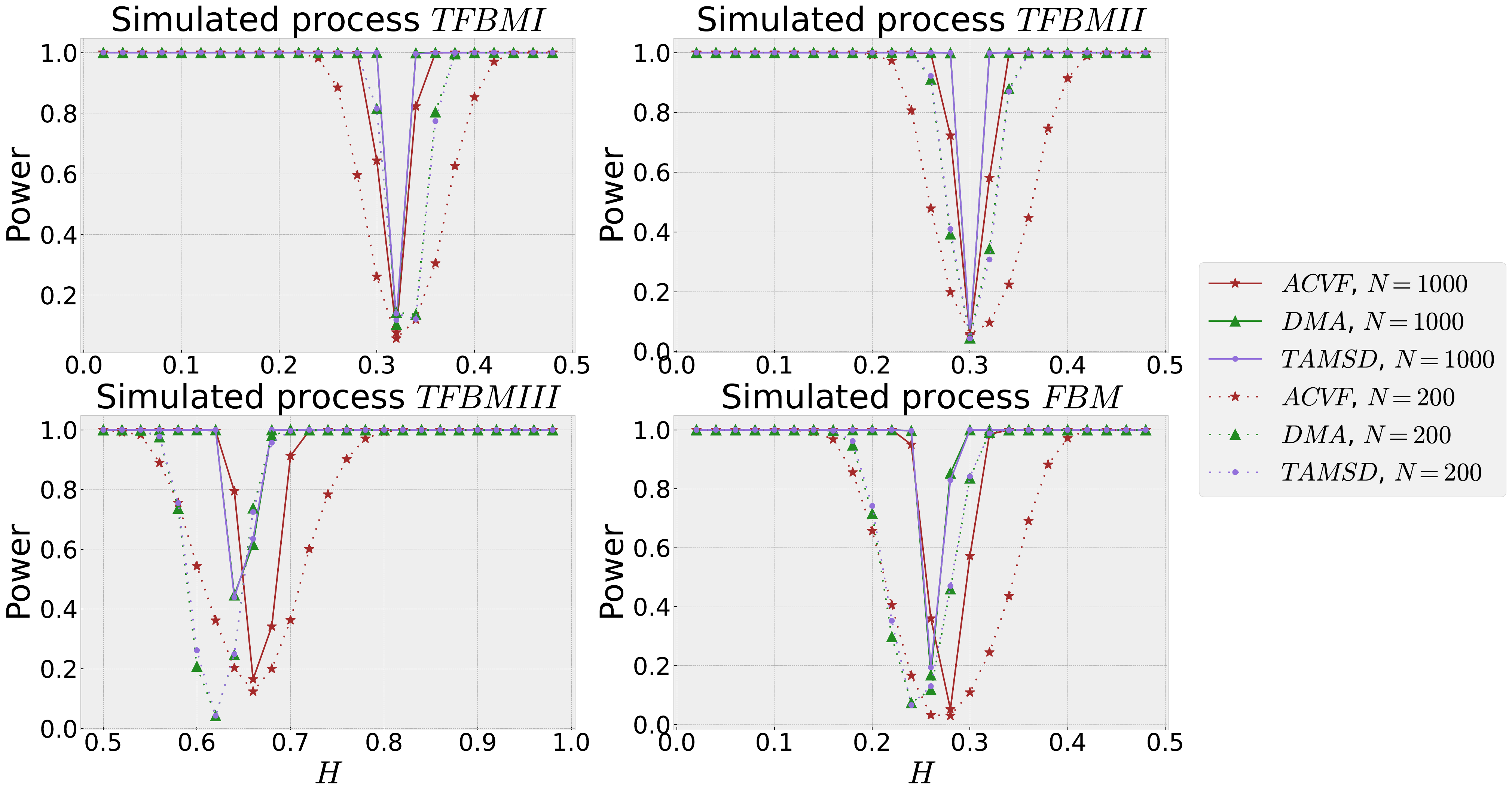}
    \caption{TFBMI\!I test power for $H_0 = 0.3$ and $\lambda_0  = 0.3$ calculated for TFBMI, TFBMI\!I, TFBMI\!I\!I and FBM trajectories with different $H$ values for two sample lengths ($N \in \{200, 1000\}$).}
    \label{fig:tfbm2_power_var_H_H1_l1}
\end{figure}
\begin{figure} [!h]
    \centering
    \includegraphics[width=0.99\linewidth]{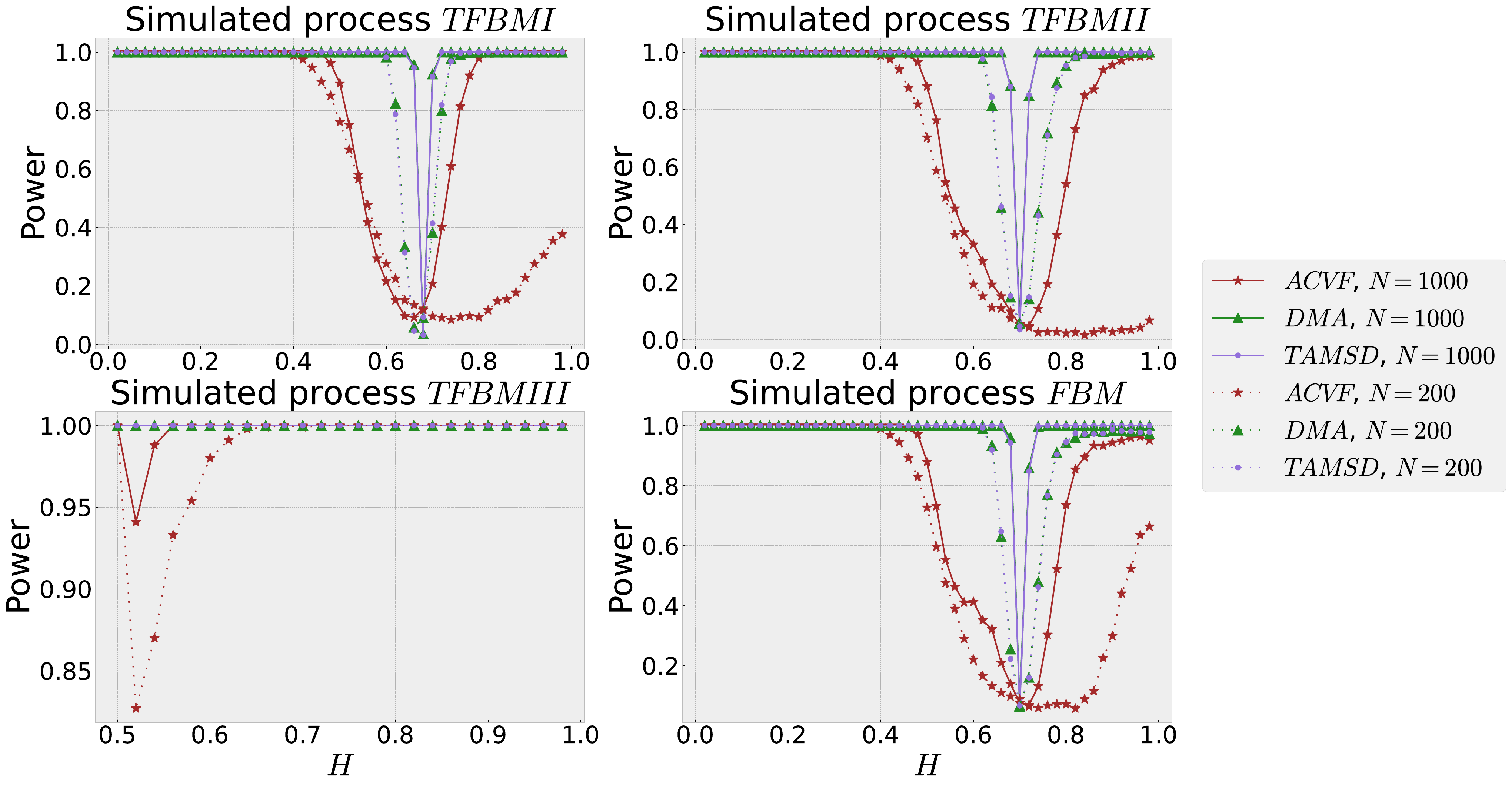}
    \caption{TFBMI\!I test power for $H_0 = 0.7$ and $\lambda_0 = 0.3$ calculated for TFBMI, TFBMI\!I, TFBMI\!I\!I and FBM trajectories with different $H$ values for two sample lengths ($N \in \{200, 1000\}$).}
    \label{fig:tfbm2_power_var_H_H2_l1}
\end{figure}

The results shown in Figures \ref{fig:tfbm2_power_var_H_H1_l1}  and \ref{fig:tfbm2_power_var_H_H2_l1} are similar to those for the TFBMI. We can see again that the ACF based test seems to be the worst in avoiding type II error, which is especially visible for $H_0 = 0.7$ illustrated in Figure \ref{fig:tfbm2_power_var_H_H2_l1}. Moreover, we again observe that the test is much more powerful for longer samples.

We can observe in the right top panel of Figure \ref{fig:tfbm2_power_var_H_H1_l1} that the power of the tests quickly increases to 1 when $H$ of TFMI\!I is departing from the true value $H_0=0.3$.
When the alternatives are different processes (other panels of Figure \ref{fig:tfbm2_power_var_H_H1_l1}), we
can observe a similar behaviour, only for TFBMI the lowest power (equal to the significance level of the test) is reached for $H$ slightly larger than $0.3$, for FBM the minimum is reached for $H$ slightly less than $0.3$ and for TFBMI\!I\!I the lowest value is for $H$ between $0.6$ and $0.7$.

In Figure \ref{fig:tfbm2_power_var_H_H2_l1}, which corresponds to $H=0.7$, the situation is similar. We only notice, as before for TFBMI, that the power is almost $1$ for the TFBMI\!I\!I alternative  (left bottom panel).  Moreover, the power of the test is the lowest under the alternatives of TFBMI and FBM for $H$ slightly lower than $0.7$.

We also note that we performed the same analyses for $\lambda_0 = 2$, but the results were very similar; therefore, we do not include them in the paper.

\begin{figure} [!h]
    \centering
    \includegraphics[width=0.99\linewidth]{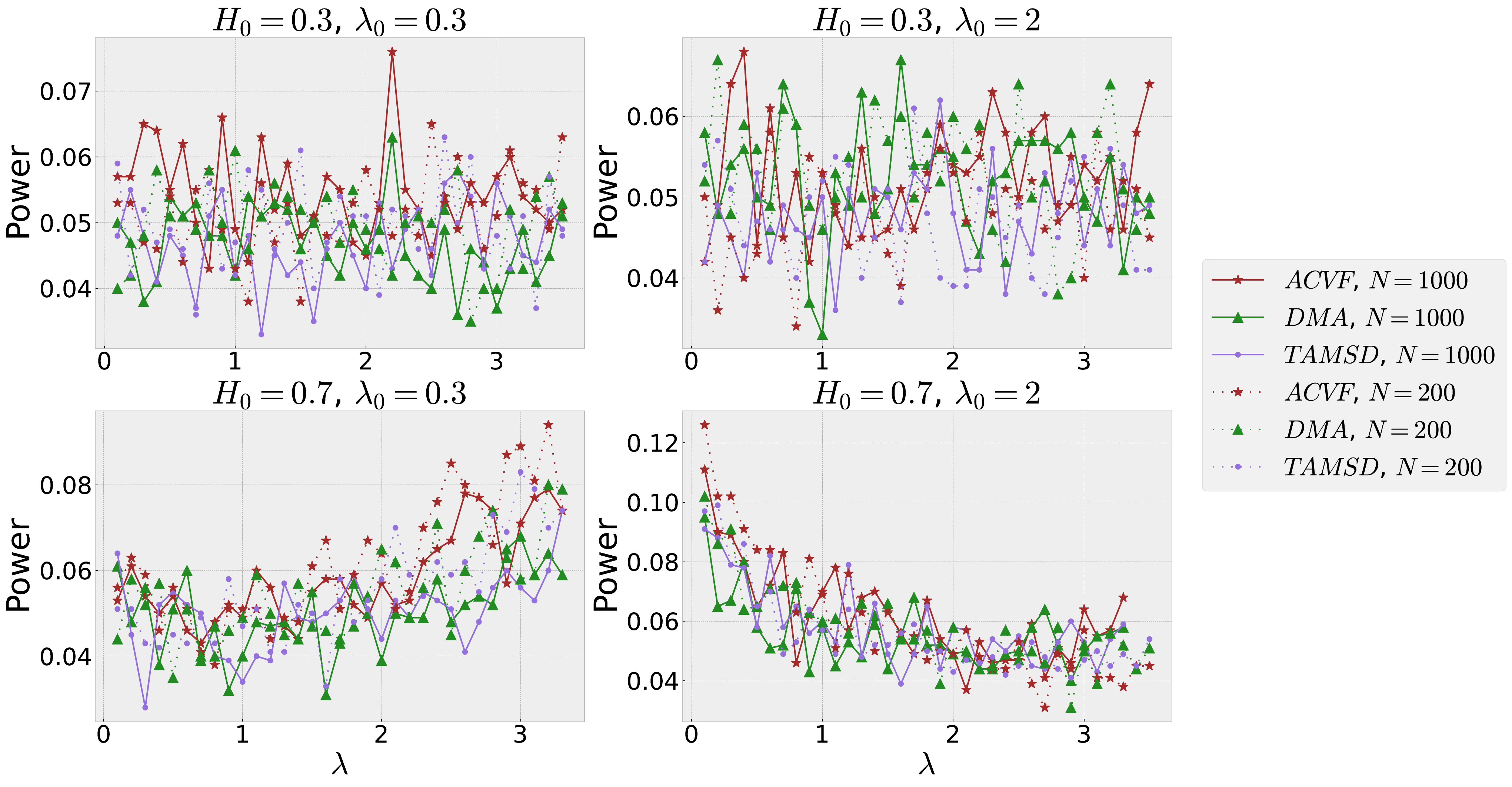}
    \caption{TFBMI\!I test power for $H_0 \in \{0.3,0.7\}$ and $\lambda_0 \in \{0.3, 2\}$  calculated for TFBMI\!I trajectories with different $\lambda$ values for two sample lengths ($N \in \{200, 1000\}$).}
    \label{fig:tfbm2_power_var_l}
\end{figure}

Finally, we examine the power of the test for fixed $H$ and varying tempering parameter $\lambda$, see Figure \ref{fig:tfbm2_power_var_l}. This time we analyse it only for the samples from the TFBMI\!I. In Figure \ref{fig:tfbm2_power_var_l} we can see that the power of the test for all alternatives is quite low around the significance level. Hence, in practice, it is difficult to reject the incorrect $\mathcal{H}_0$ hypothesis. We also observe that for higher $H_0 =0.7$ the test powers are slightly higher than for $H_0 =0.3$. 

\subsection{TFBMI\!I\!I}
\label{sec:tfbm3_power_results}
\vskip 0.3cm

The last analysed process is TFBMI\!I\!I. As before, for alternatives we use all three types of TFBM and FBM with varying Hurst exponent and tempering parameter. However, this time we use $H_0 \in \{0.9,0.7\}$, due to the constraint for the Hurst index.

\begin{figure} [!h]
    \centering
    \includegraphics[width=0.99\linewidth]{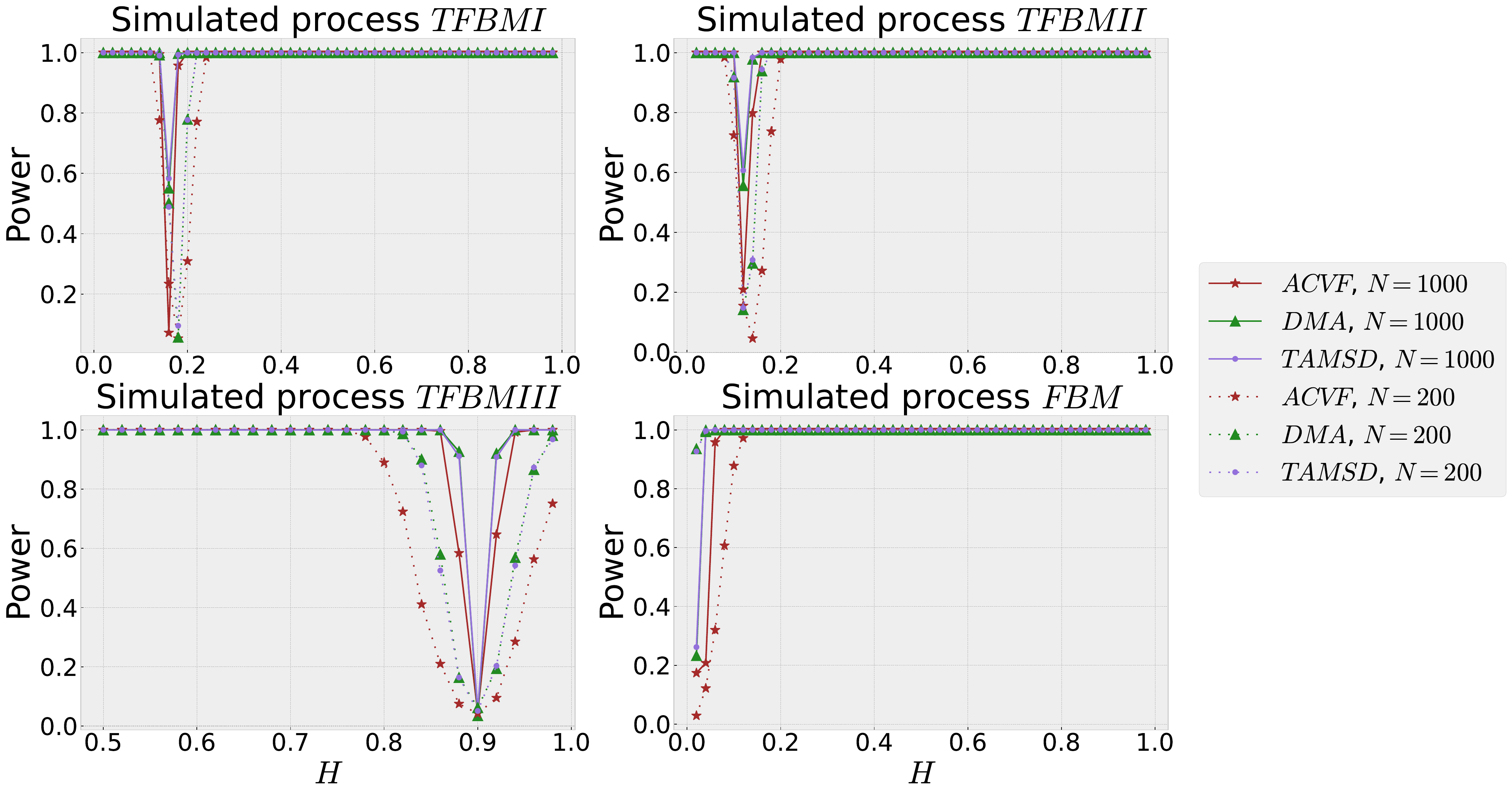}
    \caption{TFBMI\!I\!I test power for $H_0 = 0.9$ and $\lambda_0 = 0.3$ calculated for TFBMI, TFBMI\!I, TFBMI\!I\!I and $FBM$ trajectories with different $H$ values for two sample lengths ($N \in \{200, 1000\}$).}
    \label{fig:tfbm3_power_var_H_H1_l1}
\end{figure}

\begin{figure} [!h]
    \centering
    \includegraphics[width=0.99\linewidth]{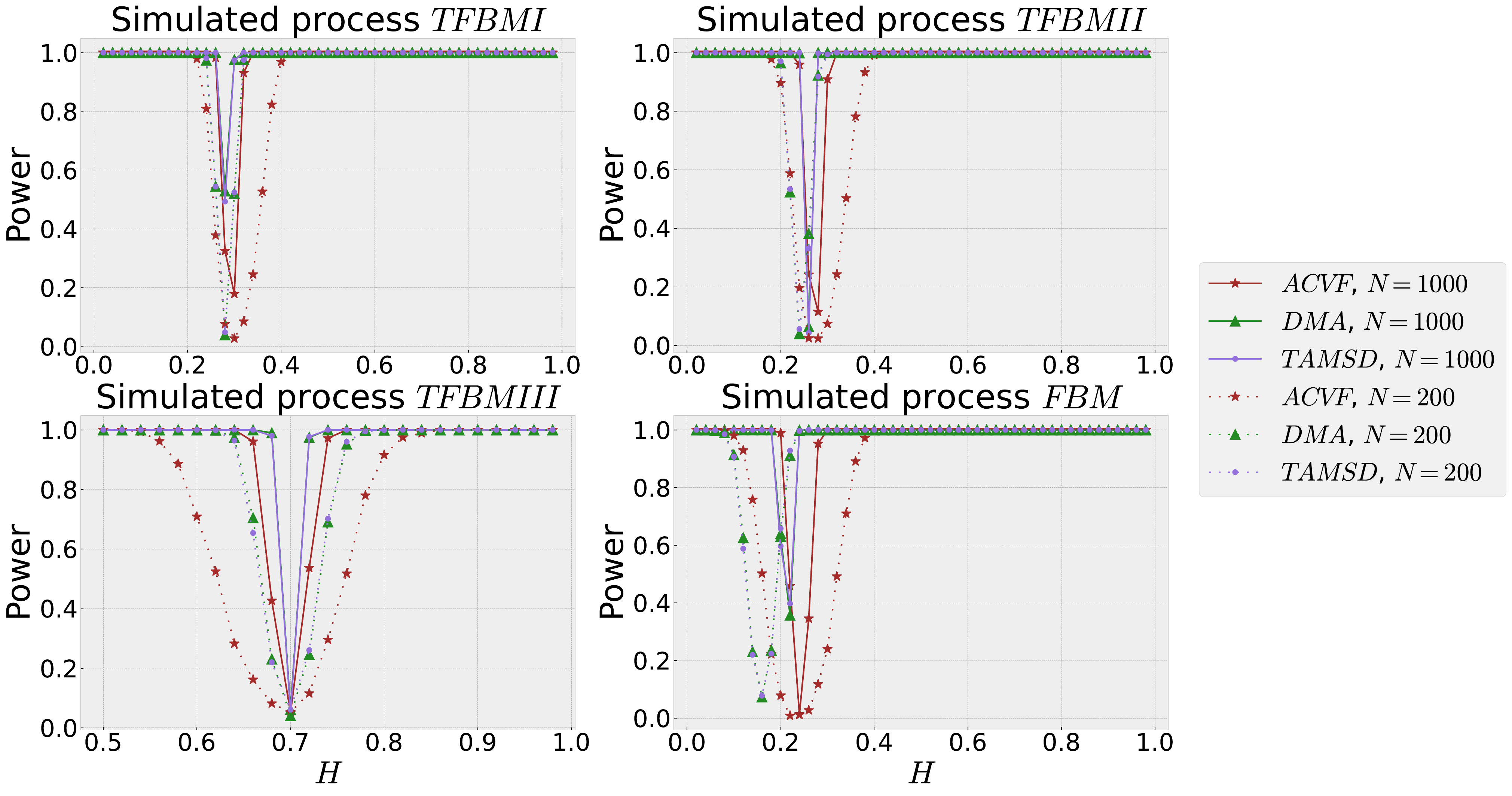}
    \caption{TFBMI\!I\!I test power for $H_0 = 0.7$ and $\lambda_0 = 0.3$ calculated for TFBMI, TFBMI\!I, TFBMI\!I\!I and FBM trajectories with different $H$ values for two sample lengths ($N \in \{200, 1000\}$).}
    \label{fig:tfbm3_power_var_H_H2_l1}
\end{figure}

In Figures \ref{fig:tfbm3_power_var_H_H1_l1} and \ref{fig:tfbm3_power_var_H_H2_l1} we can see the test powers. In the left bottom of Figure \ref{fig:tfbm3_power_var_H_H1_l1} panels we can again observe that the ACF test produces smaller powers. For other cases, where the alternatives are different processes, the power quickly increases to $1$ when $H$ deviates from the values slightly less than 0.2 for TFBMI and TFBMI\!I, and the values close to 0 for FBM.
Figure \ref{fig:tfbm3_power_var_H_H2_l1} produces similar results, only the $H$ values for which the processes are indistinguishable differ, namely for
TFBMI, TFBMI\!I and FBM, the power of the test is the lowest for $H$ slightly less than  $0.3$.

Just as for the other two types of TFBM, we now inspect if, using the introduced tests, we can differentiate between TFBMI\!I\!I samples with different values of the tempering parameter.
\begin{figure} [!h]
    \centering
    \includegraphics[width=0.99\linewidth]{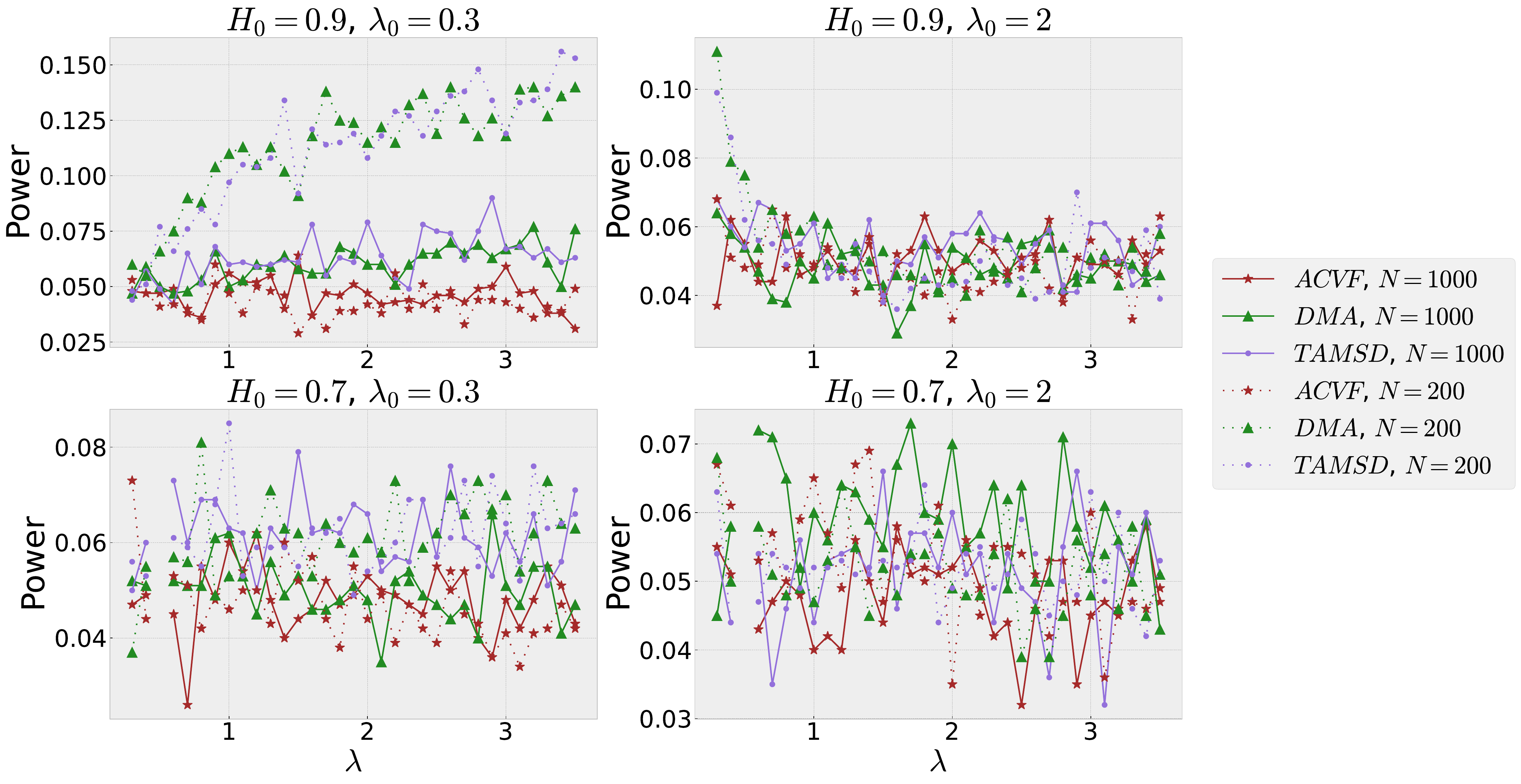}
    \caption{TFBMI\!I\!I test power for $H_0 \in \{0.9,0.7\}$ and $\lambda_0 \in \{0.3, 2\}$  generated for TFBMI, TFBMI\!I and TFBMI\!I\!I trajectories with different $\lambda$ values for two sample lengths ($N \in \{200, 1000\}$).}
    \label{fig:tfbm3_power_var_l_H1}
\end{figure}
In Figure \ref{fig:tfbm3_power_var_l_H1} we can see that the power values oscillate around $0.05$ without significant differences for various $\lambda$ and $H = H_0$ values. This indicates that for this type of TFBM, the differentiation between samples with different $\lambda$ values is even more difficult than for other types of TFBM.

\section{Conclusions}
\label{sec:conclusions}
\vskip 0.3cm

Tempered fractional Brownian motions undoubtedly provide a new framework for the description of different real phenomena. They are related to the notions of semi-long range dependence and transient anomalous dynamics \cite{MolSan,transient2022}. In this paper we concentrated on three types of TFBMs recently introduced in the literature. We note that the TFBMI\!I\!I is very diffferent from other analysed types of tempered processes. Its construction is based on tempering of the autocovariance function not the kernel of the integral representation.

Applications in wind speed modelling or single-particle tracking (SPT) experiments in biological cells have already been presented in the literature \cite{meerschaert2013tempered,MolSan}. We also note about discrete counterparts of the TFBMs, namely autoregressive tempered fractionally integrated moving average (ARTFIMA) processes, which were found to be useful in modelling of solar flare activity and SPT data \cite{solarflare2021,transient2022,artfima_biol}. In the applications, the correct recognition of the type of dynamics is crucial for the analyses.

We proposed here a testing methodology with three statistics for checking the goodness of fit of three studied TFBM types. It appeared that DMA and TAMSD based tests show similar performance and for most of the cases the ACF based test led to a lower power. The tests allow to distinguish between the tempered processes with different $H$ parameters. We also found that FBMI\!I\!I, which is defined for $H>0.5$, can be statistically indistinguishable from other TFBMs and FBM with $H<0.5$.

We found inefficacy of the introduced tests for
recognizing among tempered and not tempered FBM and also between TFBMI and TFBMI\!I,
while the tests worked well for TFBMI\!I\!I. This shows that TFBI\!I\!I is a very unique model and can be distinguished from other tempered processes using the proposed procedure.
We also showed that the tests were usually not sensitive to changes in the tempering parameter. The best efficiency of the test was observed for TFBMI for the Hurst parameter greater than 0.5. We believe that a test based on the spectral density can be a better choice to address those issues. This will be a topic of further research.

\vskip 0.5cm

\centerline{\bf\large References}

\bigskip

\Koniec
\end{document}